\def\year{2020}\relax
\documentclass[sigconf]{acmart}

\usepackage{graphicx}
\usepackage{caption}
\usepackage{subcaption}
\usepackage{booktabs}
\AtBeginDocument{%
  \providecommand\BibTeX{{%
    \normalfont B\kern-0.5em{\scshape i\kern-0.25em b}\kern-0.8em\TeX}}}
    
\copyrightyear{2020}
\acmYear{2020}
\setcopyright{none} 
\acmConference[WWW '20]{Proceedings of The Web Conference 2020}{April 20--24, 2020}{Taipei, Taiwan}
\acmBooktitle{Proceedings of The Web Conference 2020 (WWW '20), April 20--24, 2020, Taipei, Taiwan}
\acmPrice{}
\acmDOI{10.1145/3366423.3380078}
\acmISBN{978-1-4503-7023-3/20/04}

\begin{document}

\title[An Empirical Study of Android Security Bulletins in Different Vendors]{An Empirical Study of Android Security Bulletins\\in Different Vendors}


\author{Sadegh Farhang}
\affiliation{%
  \institution{Pennsylvania State University}
}
\email{smf5604@psu.edu}

\author{Mehmet Bahadir Kirdan}
\affiliation{%
  \institution{Technical University of Munich}
  }
\email{bahadir.kirdan@tum.de}

\author{Aron Laszka}
\affiliation{%
  \institution{University of Houston}
}
\email{alaszka@uh.edu}

\author{Jens Grossklags}
\affiliation{%
 \institution{Technical University of Munich}
}
\email{jens.grossklags@in.tum.de}






\begin{abstract}
Mobile devices encroach on almost every part of our lives, including work and leisure, and contain a wealth of personal and sensitive information. It is, therefore, imperative that these devices uphold high security standards. A key aspect is the security of the underlying operating system. In particular, Android plays a critical role due to being the most dominant platform in the mobile ecosystem with more than one billion active devices and due to its openness, which allows vendors to adopt and customize it. Similar to other platforms, Android maintains security by providing monthly security patches and announcing them via the Android security bulletin. To absorb this information successfully across the Android ecosystem, impeccable coordination by many different vendors is required. 

In this paper, we perform a comprehensive study of 3,171 Android-related vulnerabilities and study to which degree they are reflected in the Android security bulletin, as well as in the security bulletins of three leading vendors: Samsung, LG, and Huawei. In our analysis, we focus on the metadata of these security bulletins (e.g., timing, affected layers, severity, and CWE data) to better understand the similarities and differences among vendors. We find that (i) the studied vendors in the Android ecosystem have adopted different structures for vulnerability reporting, (ii) vendors are less likely to react with delay for CVEs with Android Git repository references, (iii) vendors handle Qualcomm-related CVEs differently from the rest of external layer CVEs. 
\end{abstract}

\keywords{Security, Android Security Bulletins, Technology Policy}


\maketitle

\allowdisplaybreaks

\section{Introduction}
\label{sec:Intro}

Nowadays, smartphones are an indispensable part of our lives, and they are supersaturated with sensitive and personal information. As a result, high levels of security are crucial in the smartphone ecosystem. Android is the dominant operating system (OS) in the smartphone ecosystem with more than one billion active devices~\cite{Android_Market_Share}. Android is released under an open-source license by the Android Open Source Project (AOSP). Due to the openness of the platform, many vendors and carriers adopted Android as their underlying platform. Within the Android ecosystem, Samsung, LG, and Huawei play important roles. In September 2019, Samsung's market share was $31.2\%$ of mobile devices worldwide, while Huawei's and LG's market shares were $10.0\%$ and $2.5\%$, respectively~\cite{Device_Market_Share}. 

Since August 2015, AOSP maintains the security of the Android platform by providing monthly security patches and publishing the details of each patch in the Android security bulletin~\cite{Android_Bulletin}. Other vendors like Samsung~\cite{Samsung_Bulletin}, LG~\cite{LG_Bulletin}, and Huawei~\cite{Huawei_Bulletin} subsequently also launched their own security bulletins.

In practice, when a vulnerability is found, disclosed, and patched, one can find the relevant information in a vulnerability database, the Android security bulletin, or a vendor's security bulletin. However, due to customization and differences in hardware, a vulnerability in Android is not necessarily applicable to all vendors and devices~\cite{farhang2018economic, wu2013impact, aafer2016harvesting}. Therefore, one has to search for a vulnerability in a vendor's security bulletin to determine whether it is applicable to a vendor's device. However, even if a vendor has a security bulletin, it is possible that a vulnerability has not been mentioned in the bulletin yet, but it may appear later. Misinformation or delays in different vendors' security bulletins could mislead security practitioners. Moreover, the absence of relevant information in a vendor's security bulletin is an important indicator that millions of smartphones may be unpatched and vulnerable. 

In this paper, to the best of our knowledge, we perform the first comprehensive study of how vendors handle Android-related vulnerabilities in their security bulletins. We collect a total of 3,171 unique CVEs from Android, Samsung, LG, and Huawei security bulletins, as well as further data from Google Git~repositories, and CVEDetails, which provides detailed information for each vulnerability. Thereby, we focus on the four vendors that regularly publish security bulletins. We shed light on the security practices in the broader Android ecosystem and, specifically, how vendors handle Android-related vulnerabilities. In summary, our paper makes the following key contributions:

\begin{itemize}
    \item We show that each vendor adopted a different approach for announcing CVEs in its security bulletins. Samsung is the only vendor mentioning CVEs that are not applicable to its devices. Therefore, the majority of CVEs originating from Android security bulletins appeared in Samsung's bulletins ($99.55\%$). In contrast, for LG and Huawei, the ratio of explicitly mentioned vulnerabilities is only $78.16\%$ and $52.61\%$, respectively, creating significant uncertainty.
    
    \item In terms of delay between Android security bulletins and a vendor's security bulletins, we find that Huawei does not have any time differences for $97.0\%$ of its CVEs. In contrast, Samsung and LG do not have any time difference for only $44.7\%$ and $39.44\%$ of their CVEs, respectively.
    
    \item We find that there is no delay for almost all CVEs that have an AOSP Git repository reference in Android security bulletins, which is true for all vendors.
    
    \item Time differences among vendors appear mostly for CVEs of the \textbf{external} and \textbf{kernel} Android OS layers. Moreover, with respect to the average delay in the external layer, we find that Samsung and LG handle Qualcomm-related CVEs with longer delay than the rest of the external layer CVEs. In contrast, for Huawei the average delay for Qualcomm-related CVEs is lower than the rest of external layer CVEs. 
\end{itemize}

\section{Data Collection}
\label{sec:Data}
Since each vendor publishes its vulnerability patches in its own security bulletin, each of them has its unique format and set of fields for describing vulnerabilities. Moreover, none of these vendors provide data in a standard, machine-readable format, such as JSON or XML. Therefore, we  built a designated crawler and content parser for each vendor. To crawl the vendors' websites, we used Selenium Browser Automation~\cite{Selenium}. We collected data  until August 2019.

Android and Huawei security bulletins include only CVEs, while Samsung and LG security bulletins contain not only CVEs but also their unique vulnerability identifiers: LG Vulnerabilities and Exposures (LVE) and Samsung Vulnerabilities and Exposures (SVE). 

On Android security bulletins, we scraped all CVEs from August 2015 (i.e., first published bulletin) until August 2019. Early versions of the Android security bulletin have different field names than the most recent version. For instance, the field \textit{Updated AOSP Versions} has different names, such as \textit{Affected Versions} in August 2015~\cite{Android_Security_Bulletin_2015_08} and \textit{Updated Versions} in December 2015~\cite{Android_Security_Bulletin_2015_12}. As a result, we needed different crawlers and content parsers even for a specific vendor's security bulletins. 

CVEs on Android security bulletins might also contain a field called \textit{References} linking to the Android AOSP Git Repository~\cite{google_aosp_git}, which shows all of the commit details. When a vulnerability patch has a reference field, we also scraped the commit details of that particular vulnerability patch. 

We scraped LG and Samsung CVEs beginning from the start date of their security bulletins, May 2016 and October 2015, respectively. On Huawei, however, there are two different security bulletins; we scraped both. The first one appears as \textit{Huawei EMUI/Magic UI} security updates, which started in  December 2017~\cite{huawei_emui}. The second one is called \textit{Security Advisories}~\cite{huawei_security_advisory}, which started in 2012. On these security advisories, there are also CVE IDs that reference the particular security advisory. However, not all of these security advisories are accompanied by the corresponding CVE. In total, 817 unique CVEs have been mentioned in these security advisories and only two of them are common with the first Huawei bulletins.  Moreover, if we consider the official start date of Huawei security bulletins in December 2017, Huawei security advisory has only 6 common CVEs with the Android security bulletins, which does not impact our results. Here, we mainly focus on the difference between Android security bulletins and a vendor's security bulletin. Hence, we do not consider the Huawei security advisory in our analysis.

\begin{figure}[h]
\centering
\includegraphics[width=0.3\textwidth]{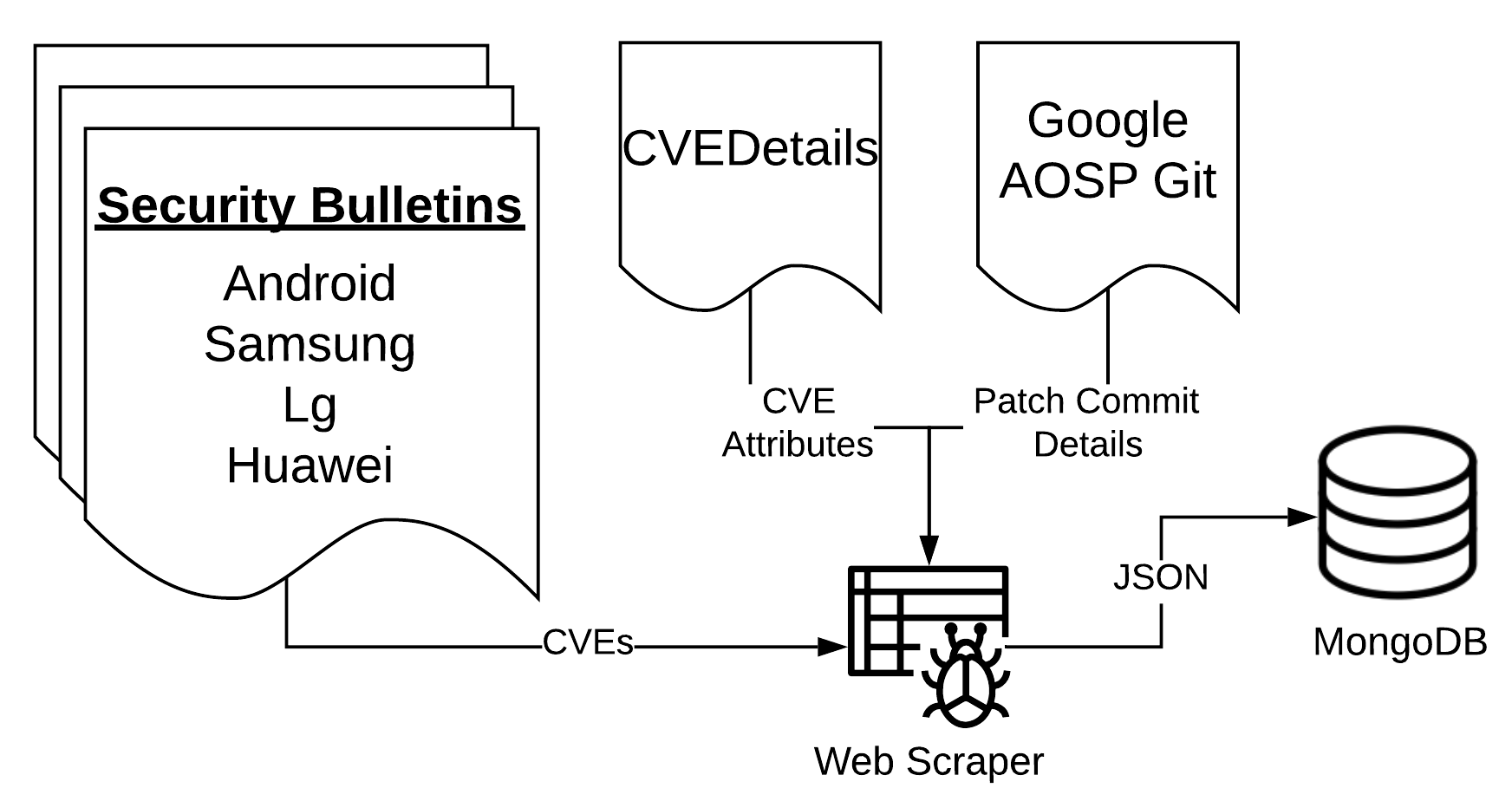}
\caption{Data scraping schema.}
\label{fig:data_scraping_schema}
\end{figure}

Aside from these security bulletins, we also scraped CVEDetails~\cite{Cve_Details} to gain additional attributes of the vulnerabilities. After scraping all of these vulnerabilities, we converted them to a common JSON~\cite{JSON} format and stored them in MongoDB~\cite{MongoDB}. 
Figure~\ref{fig:data_scraping_schema} shows the overall data collection process described above.\footnote{The dataset is available at \url{https://github.com/culture67/Android-Bulletin-Data}} 

\vspace{-2mm}
\textbf{Limitations:}
\label{sub:limit}
Our work has the following limitations. \textit{First}, we focus only on Android security bulletins and on the three Android vendors that, to our knowledge, have established comprehensive security bulletins. \textit{Second}, for each vendor, our analysis is limited to the time since the vendor started publishing its security bulletins. We cannot claim our results are valid from the Android commercialization date (i.e., 2008) til the start of a vendor's security bulletins. But, our analysis is representative of the current ecosystem rather than the past. \textit{Third}, the available information for each CVE varies. As an example, for some CVEs, we have a reference link; but for some, we do not have any. We use this type of information plus component and category names in Android security bulletins to perform our Android stack layer analysis. Hence, our analysis for Android stack layers is limited to those CVEs for which we have the corresponding information. \textit{Fourth}, we are aware that many other factors can affect how a vendor manages its security bulletins or how a vendor mentions and describes them. In our analysis, we limit ourselves to only a vendor's security bulletins and report what we observe from these. As such, we defer code analysis to future work, and instead focus on a high-level analysis of the security bulletins. 

\section{Results}
\label{sec:Results}
Next, we analyze the data that we have collected to understand how Android vendors manage their security bulletins. 

\subsection{Data Characterization}
\label{sub:characterize}

In total, we scraped 3,171 unique CVEs from the four different vendors' security bulletins (see Table~\ref{tab:No_of_CVE}). 

\begin{table}[h]
    \centering
    \begin{tabular}{|c||c|c|c|c|} \hline
         & Android & Samsung & LG & Huawei \\ \hline\hline
        No. of CVEs & 2,705 & 2,587 & 2,023 & 816\\ \hline
        Unique to & 202 & 157 & 12 & 31\\ \hline
        Launch Date & Aug. 2015 & Oct. 2015 & May 2016 & Dec. 2017 \\ \hline
    \end{tabular}
    \caption{Number of CVEs and security-bulletin launch dates for different vendors. Row ``Unique to'' shows the number of CVEs that are  mentioned only in that vendor's security bulletins.}
    \label{tab:No_of_CVE}
\end{table}


\subsection{Bulletin Management}
\label{sub:Parts}

\begin{figure*}
     \centering
     \begin{subfigure}[b]{0.28\textwidth}
         \centering
         \includegraphics[width=\textwidth]{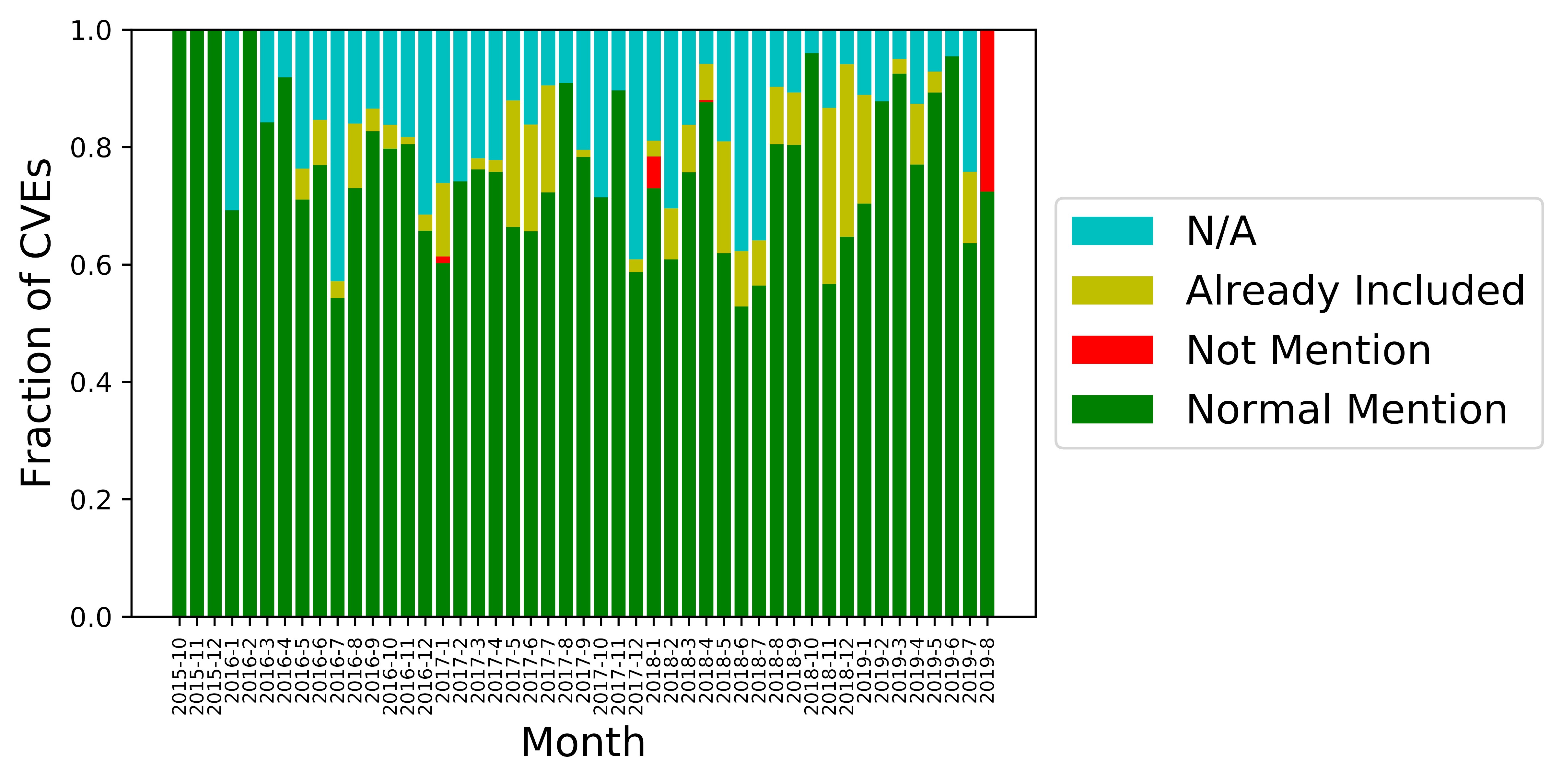}
         \caption{Samsung}
         \label{fig:Sam_manage}
     \end{subfigure}
     \hfill
     \begin{subfigure}[b]{0.28\textwidth}
         \centering
         \includegraphics[width=\textwidth]{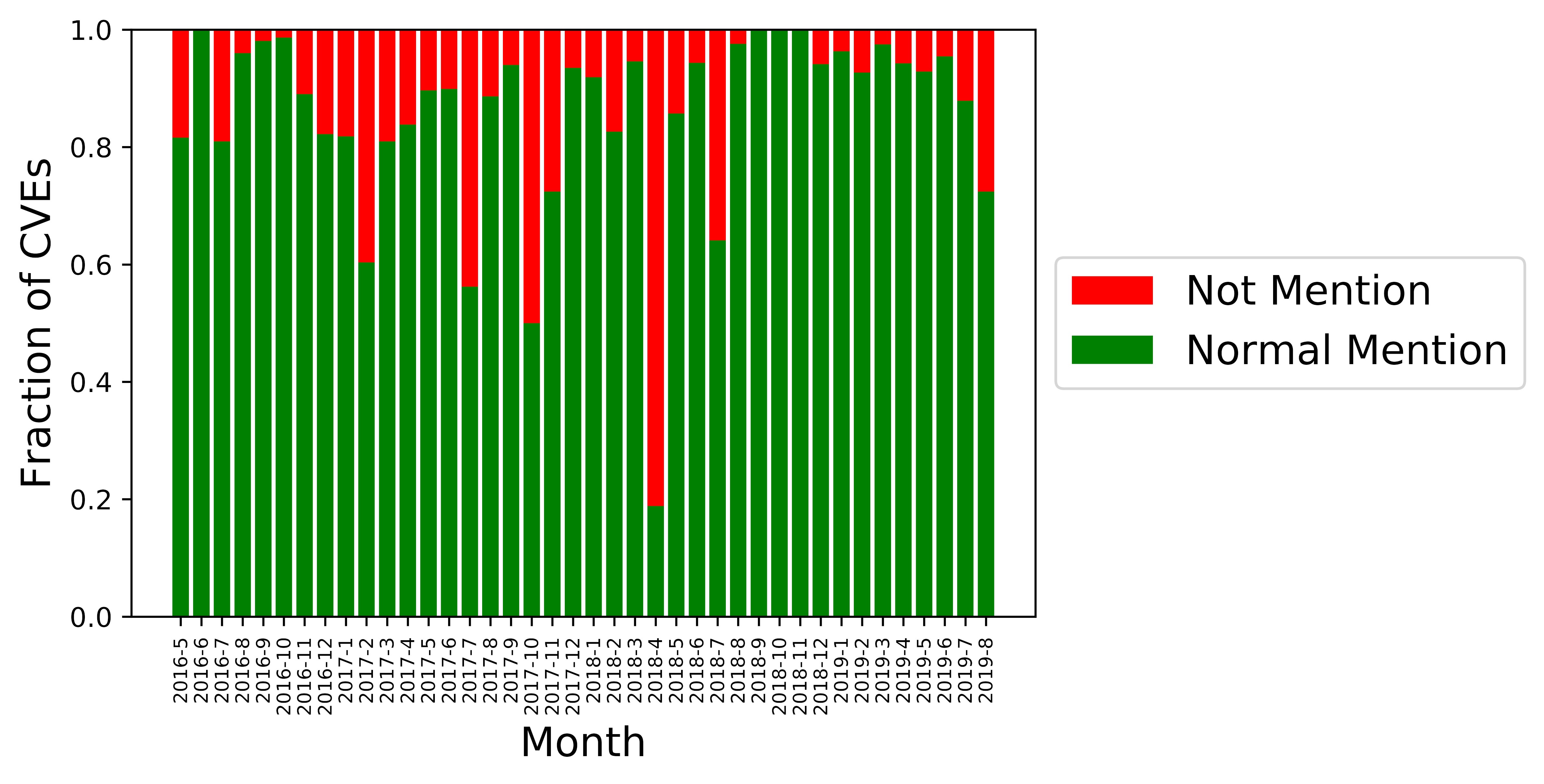}
         \caption{LG}
         \label{fig:LG_manage}
     \end{subfigure}
     \hfill
     \begin{subfigure}[b]{0.28\textwidth}
         \centering
         \includegraphics[width=\textwidth]{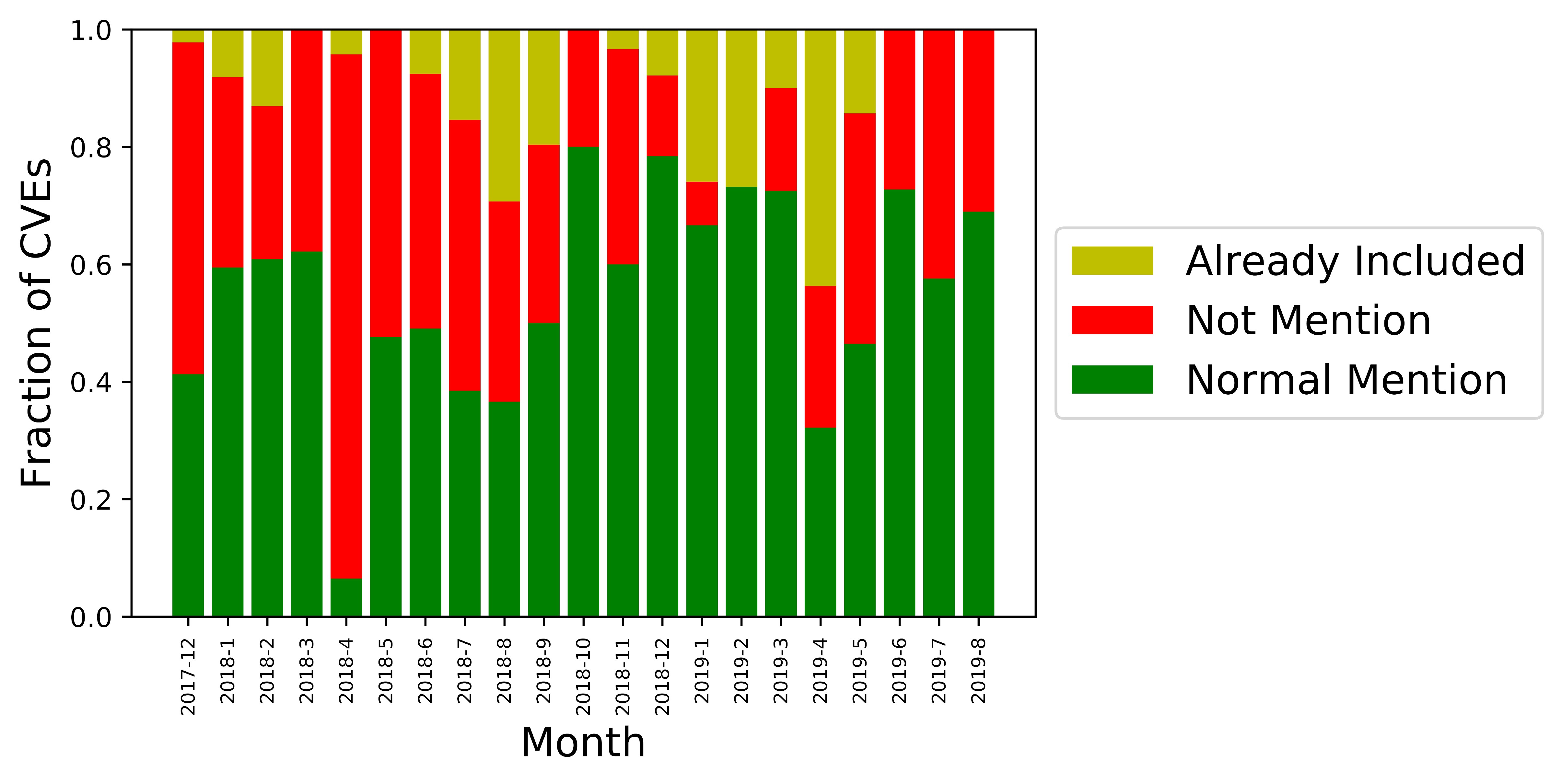}
         \caption{Huawei}
         \label{fig:Huawei_manage}
     \end{subfigure}
        \caption{Vendors' management of security bulletins compared to Android. }
        \label{fig:mange}
\end{figure*}

Different vendors have adopted different approaches for handling their security bulletins. LG announces CVEs with their corresponding severity levels monthly. Huawei employs the same practice, but it also has a part for CVEs that are \textit{already included in previous updates}\footnote{We use \textit{already included} hereafter.} without mentioning the exact date of bulletin. In addition to severity levels and already included CVEs, Samsung utilizes the labels \textit{Not applicable to Samsung devices}\footnote{We use \textit{N/A} or \textit{not applicable} hereafter.} and \textit{in addition} (which only has been used by Samsung only once, in February 2018).

Figure~\ref{fig:mange} summarizes how these three vendors announce CVEs in their security bulletins and includes a comparison to Android. In this figure, for each month, we consider the CVEs that have been announced on the Android security bulletin of that month. CVEs mentioned in the bulletins with a \textit{severity level} or \textit{in addition} are considered as \textit{Normal Mention}. \textit{Already Included} and \textit{N/A} represent CVEs mentioned in the already included part and not applicable part of a vendor's security bulletin, respectively. \textit{Not Mentioned} represents those CVEs that have not been mentioned at a vendor's security bulletin so far compared to the Android security bulletins.

According to Figure~\ref{fig:mange}, Samsung has mentioned most of the CVEs ($99.55\%$), that previously appeared in the Android security bulletins, which stands in contrast to LG and Huawei ($78.16\%$ and  $52.61\%$, respectively). As we can see in Figure~\ref{fig:Sam_manage}, there is a rise in \textit{not mentioned} CVEs in August 2019. One likely explanation is that Samsung will announce them in the upcoming months beyond the date we gathered our data. Moreover, in April 2018, there are many CVEs that have not been mentioned in both LG and Huawei contrary to Samsung. The reason leading to this is how Android announces Qualcomm related CVEs. In that month, the Android security bulletin had a section for 225 cumulative updates for Qualcomm components to associate them with a patch level. These CVEs were shared by Qualcomm with their partners between 2014 and 2016. Samsung mentioned all these CVEs two months earlier in its security bulletin as  \textit{in addition} (the only time Samsung has used this label so far).   

Samsung started using the label \textit{not applicable} in January 2016. 
Now, there are 537 CVEs associated with this label. For example, from the beginning of LG security bulletins, i.e., May 2016, Samsung announced 527 CVEs with a not applicable label. From these 527, 265 have not been mentioned in the LG security bulletins so far. From December 2017 (Huawei bulletin's start date), there are 233 such CVEs and 159 of them have not been mentioned in Huawei's security bulletins, yet. Due to vendors' customization practices, it is expected that some CVEs are not applicable for all vendors. Nonetheless, the difference is surprising.
For Samsung, due to the not applicable label, we have some assurance that these CVEs indeed are not relevant. But, for LG and Huawei, we are left with a large degree of uncertainty. Therefore, a key reporting suggestion is that all vendors should introduce a section or label for not applicable CVEs (and references to their devices). 


\subsection{Time Comparison and Android Layers}

\begin{figure}[h]
     \centering
     \begin{subfigure}[b]{0.21\textwidth}
         \centering
         \includegraphics[width=\textwidth]{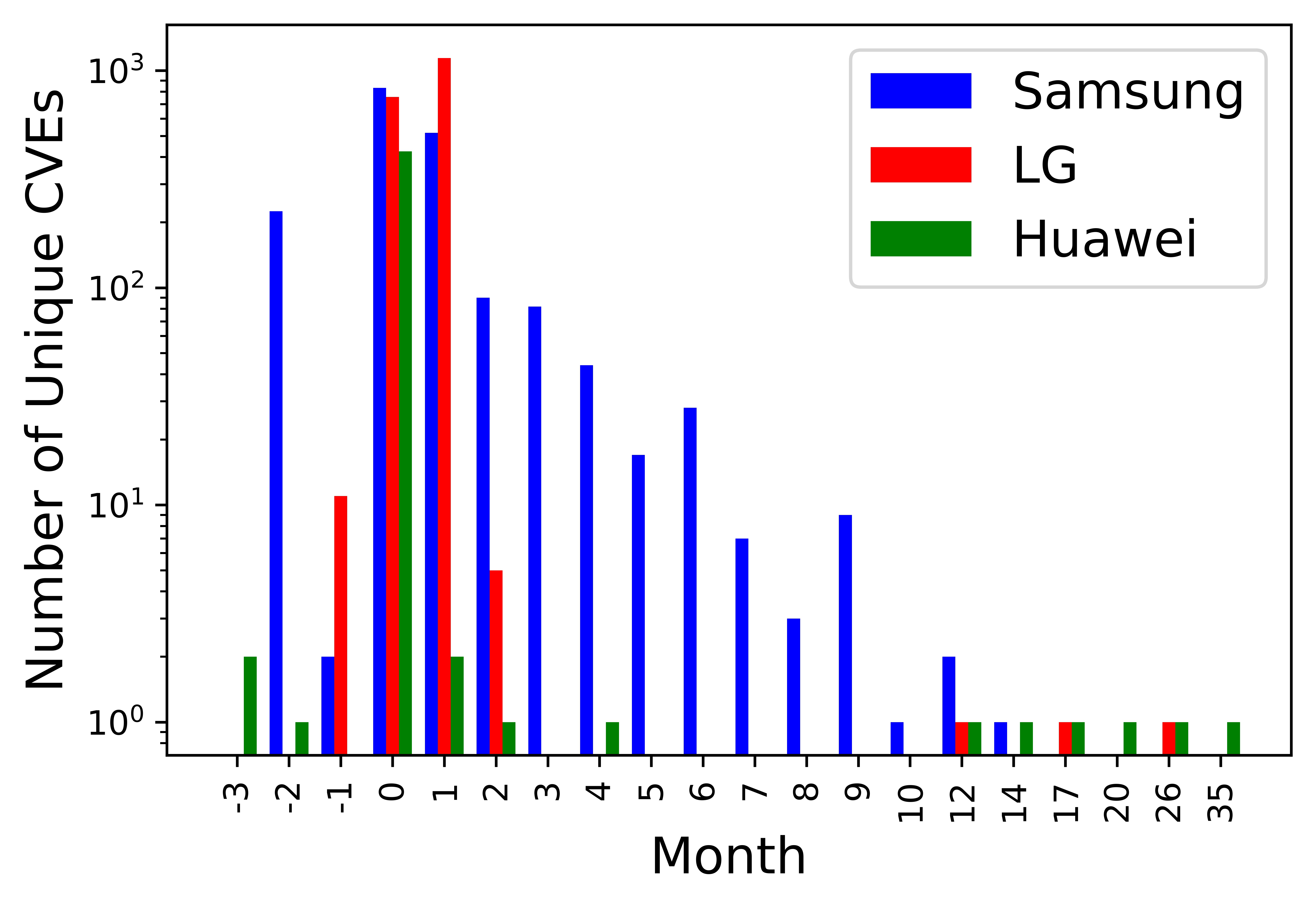}
         \caption{Absolute Number}
         \label{fig:delay_all}
     \end{subfigure}
     \hfill
     \begin{subfigure}[b]{0.21\textwidth}
         \centering
         \includegraphics[width=\textwidth]{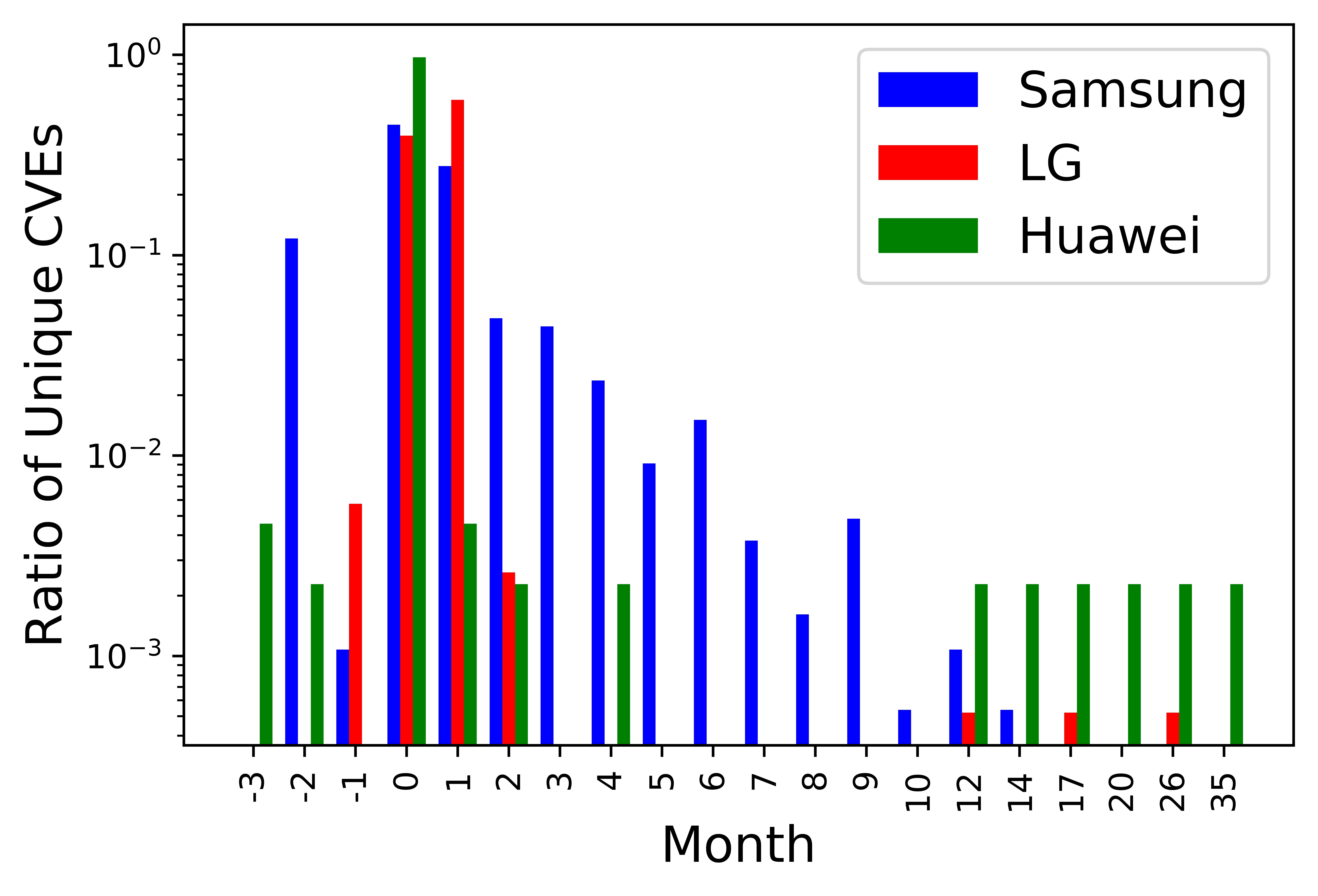}
         \caption{Ratio}
         \label{fig:delay_percent}
     \end{subfigure}
     \hfill
     \caption{Distribution of vendors' time difference from Android security bulletin.}
        \label{fig:delay}
\end{figure}

We further investigated the timeline of normal mentions in these vendors compared to Android (see Figure~\ref{fig:delay}). Figure~\ref{fig:delay_all} represents the absolute number of CVEs for each vendor's security bulletin with the corresponding time difference from the Android security bulletin (CVEs appeared in a vendor's security bulletins once). The positive (negative) number means that a vendor is slower (faster) than Android to mention a CVE in its security bulletins. Figure~\ref{fig:delay_percent} represents the ratio of the time difference. As we see in Figure~\ref{fig:delay}, Huawei does not have any time differences with Android for $97.03\%$ of CVEs compared to Samsung and LG, $44.73\%$ and $39.44\%$, respectively. LG mentions CVEs one month after the Android security bulletin for $59.57\%$ of its CVEs, while this number is $27.80\%$ for Samsung. For Samsung, there exists a considerable number of CVEs that are mentioned in Samsung security bulletins two months earlier than Android. Note that all these CVEs are those that have been mentioned with the label \textit{in addition} in February 2018. 

It is useful to further break down vendors' behaviors in terms of Android stack layers to better understand the source of differences among vendors. To achieve that, we first need to find the corresponding layer for a CVE since this information is not publicly available in security bulletins. In doing so, we use three attributes (which is discussed in detail in our previous work~\cite{farhang2019hey}), \textit{component name} (a column for some CVEs in Android security bulletins), \textit{category name} (in an Android security bulletin, CVEs appear under specific category), and the combination of a branch path with the changed file patch for CVEs that have an Android Git repository. These three attributes are not available for all CVEs. As a result, we  can only find layer information for some CVEs and remove the rest from our analysis. Only two CVEs relate to the Android runtime layer. Therefore, we exclude them from our analysis. In the following, we consider each layer separately. 

\textbf{External Layer.} This layer has 1772 unique CVEs in total. Samsung has 1761 (1216 Qualcomm-related) CVEs in the external layer in which 80 and 5 of them have two or three mentions in the Samsung security bulletins, respectively.  The majority of CVEs with N/A are in this layer (357 out of 537) and Samsung mentions them within one month after the Android security bulletins. LG has 1213 (821 Qualcomm-related) unique CVEs and 21 of them are mentioned twice. 291 unique CVEs (218 Qualcomm-related) belong to Huawei and 10 of them are mentioned twice in Huawei security bulletins. 

Contrary to Huawei, Samsung and LG are faster in announcing non-Qualcomm CVEs in their security bulletins than Qualcomm-related CVEs\footnote{For this analysis and the rest, we only focus on normal CVEs with one mention. We exclude already included CVEs since we do not know the exact time of the corresponding bulletin. The reasons we do not consider multiple mentions are (i) it is not common among vendors and (ii) it is not clear which delay we should consider.} (see Table~\ref{tab:external}). The difference for Huawei is not significant (Mann-Whitney test, $p = 0.103$), but the difference is significant for both Samsung (Mann-Whitney test, $p<0.0001$) and LG (Mann-Whitney test, $p<0.0001$). We also perform an analysis of variance (ANOVA) test for both Qualcomm-related CVEs ($p < 0.0001$) and the rest of external layer CVEs ($p < 0.0001$). These show that the average delay for these three vendors is significantly different for both Qualcomm-related CVEs and the rest of the external layer~CVEs. 

\begin{table}[h]
    \centering
    \footnotesize{
    \begin{tabular}{|c||c|c|c|} \hline
         & Qualcomm & Rest of External Layer & Application Layer \\ \hline\hline
        Samsung & 1.753 & 0.837 & 0.571 \\ \hline
        LG & 0.957 & 0.428 & -0.0541\\ \hline
        Huawei & -0.0588 & 0.0152 & 0 \\ \hline
    \end{tabular}}
    \caption{Average delay (in months) of CVEs in external layer and application layer.}
    \label{tab:external}
\end{table}


\textbf{Application Layer.} We find 50 unique CVEs in the application layer. All of them are mentioned in Samsung, but 9 of them are N/A to Samsung. 44 of them are mentioned once. LG mentioned 37 of them once in its security bulletins and only one of them twice. Huawei mentioned 16 CVEs of this layer once. For both Samsung and Huawei, there is no already included mention for this layer. See application column of Table~\ref{tab:external} for the average delay. LG and Huawei do not introduce any delay for CVEs in this layer.

\textbf{Application Framework Layer (AFL).} There are 128 unique CVEs in this layer. Samsung mentioned 126 of them. For LG and Huawei, we have 108 and 41, respectively. The AFL column in Table~\ref{tab:Kernel} shows the average delay of this layer.  

\begin{table}[h]
    \centering
    \footnotesize{
    \begin{tabular}{|c||c|c|c|c|} \hline
         & AFL & NL & HAL & Kernel \\ \hline\hline
        Samsung & 0.241 & 0.0766 & 0.0515 & 1.296 \\ \hline
        LG & 0.1373 & 0.139 & 0.0522 & 0.972 \\ \hline
        Huawei & 0 & 0.811 & 0 & 0.0357 \\ \hline
    \end{tabular}}
    \caption{Average delay (in months) of CVEs in application framework, native, hardware abstraction, and Kernel layers.}
    \label{tab:Kernel}
\end{table}

\textbf{Native Library (NL).} In Android, there are 266 unique CVEs. 247 of them are mentioned in the Samsung bulletins. LG has mentioned 191 of them and 45 of them are mentioned in Huawei so far. Samsung has the lowest average delay in this layer compared to other layers, i.e., 0.0766 months. On the other hand, Huawei has the highest average delay for CVEs of this layer, i.e., 0.811 (see Table~\ref{tab:Kernel}).

\textbf{Hardware Abstraction Layer (HAL).} In this layer,  we find 146 unique CVEs. 145 of them are mentioned in Samsung. 137 of them are mentioned in LG security bulletins and Huawei has mentioned 106 of them. For average delay, see Table~\ref{tab:Kernel}. 

\textbf{Kernel Layer.} We have 213 unique CVEs in the Kernel layer. Samsung has mentioned 211 of them in its security bulletins in which 43 of them are ``not applicable'' to Samsung devices. LG mentioned 186 of them and 40 of them are mentioned in Huawei. It is the only layer in which other vendors have never been mentioning any CVEs sooner than Android security bulletins. In other words, there does not exist any negative delay for CVEs of this layer.  We also perform an ANOVA test on the delay of these three vendors and the difference among them is significant ($p<0.00001$). 

\begin{table*}[]
\small{
\begin{tabular}{|c|c|c|c|c|c|c|c|c|c|}
\hline
Vendor                & \multicolumn{3}{c|}{Huawei vs. Samsung}         & \multicolumn{3}{c|}{LG vs. Huawei}        & \multicolumn{3}{c|}{Samsung vs. LG}   \\ \hline
Layer                 & H \textless S       & H = S & H \textgreater S & L \textless H & L = H & L \textgreater H & S \textless L & S = L & S \textgreater L \\ \hline\hline
External              & 71 (1.014) & 62    & 1 (1)   & 2 (1)         & 62    & 103 (1.097)      & 29 (1)        & 610   & 218 (2.95)       \\ \hline
Application           & 0                   & 14    & 0                & 0             & 15    & 0                & 0             & 29    & 3 (1.33)         \\ \hline
Application Framework & 1 (2)               & 34    & 0                & 0             & 33    & 1 (1)            & 4 (1)         & 89    & 11 (1.545)       \\ \hline
Native Library        & 1 (2)               & 38    & 0                & 0             & 38    & 1 (2)            & 1 (1)         & 180   & 6 (1)            \\ \hline
Hardware Abstraction  & 3 (1.33)            & 101   & 0                & 0             & 98    & 6 (1.667)        & 4 (1)         & 131   & 2 (1)            \\ \hline
Kernel                & 24 (1)              & 1     & 0                & 0             & 1     & 27 (1)           & 9 (1)         & 106   & 35 (1.6)         \\ \hline
\end{tabular}
\caption{Comparison of the vendors' response times. S, L, and H represent Samsung, LG, and Huawei, respectively.}\label{tab:layers}}
\end{table*}

We also compare the response time of two vendors in each layer for common CVEs between them (see Table~\ref{tab:layers}). In this table, each entry represents the number of times a vendor is faster compared to another one in announcing a CVE in its security bulletins and the number in parenthesis shows the average difference in a month. As an example, $H < S$ means that Huawei is faster than Samsung in announcing CVEs of a layer. As we can see in this table, Huawei is rarely slower than Samsung and LG in all layers. This was expected as we see Huawei mentions CVEs in its bulletins mostly without delay (see Figure~\ref{fig:delay_percent}). Furthermore, all three vendors treat CVEs from all layers almost the same except \textbf{External} and \textbf{Kernel} layers.  Huawei is faster than both Samsung and LG for $52.98\%$ and $61.68\%$ of \textit{external} layer CVEs, respectively. LG is faster than Samsung for $25.44\%$ of the external layer's CVEs. In other words, the external layer CVEs in Samsung are rarely mentioned sooner than the corresponding CVEs in Huawei and LG security bulletins. For Kernel layer CVEs, Huawei is almost always faster than both Samsung and LG. LG is faster than Samsung for $23.33\%$ of the CVEs. This shows that vendors handle CVEs originating from Kernel and External layers differently in terms of announcements in their security bulletins. Huawei is the fastest one, while Samsung is the slowest.

The above analysis also suggests that CVEs with external references/repositories like Qualcomm and Kernel may introduce some delay for vendors to announce these CVEs in their security bulletins. Further, in Android security bulletins, some CVEs have an AOSP Git repository. Hence, we investigate whether vendors handle CVEs with the AOSP Git repository differently in terms of a delay from the Android security bulletin. In total, 825 CVEs have an AOSP Git repository reference. Note that we restrict our analysis to those CVEs that are only mentioned once in a vendor's security bulletins. According to Table~\ref{tab:AOST_Git}, for most CVEs with an AOSP Git repository reference, there is no delay in a vendor's security bulletin to mention a CVE. 

\begin{table}[]
    \centering
    \begin{tabular}{|c||c|c|c|} \hline
         & Delay & Without Delay & Already Included \\ \hline\hline
        Samsung & 25 & 611 & 3\\ \hline
        Huawei & 4 & 173 & 10 \\ \hline
        LG &  24& 535& - \\ \hline
    \end{tabular}
    \caption{Number of CVEs with AOSP Git reference by delay status for different vendors.}
    \label{tab:AOST_Git}
\end{table}


\subsection{Severity Level and Vulnerability Type}

\begin{figure}[h]
     \centering
     \begin{subfigure}[b]{0.15\textwidth}
         \centering
         \includegraphics[width=\textwidth]{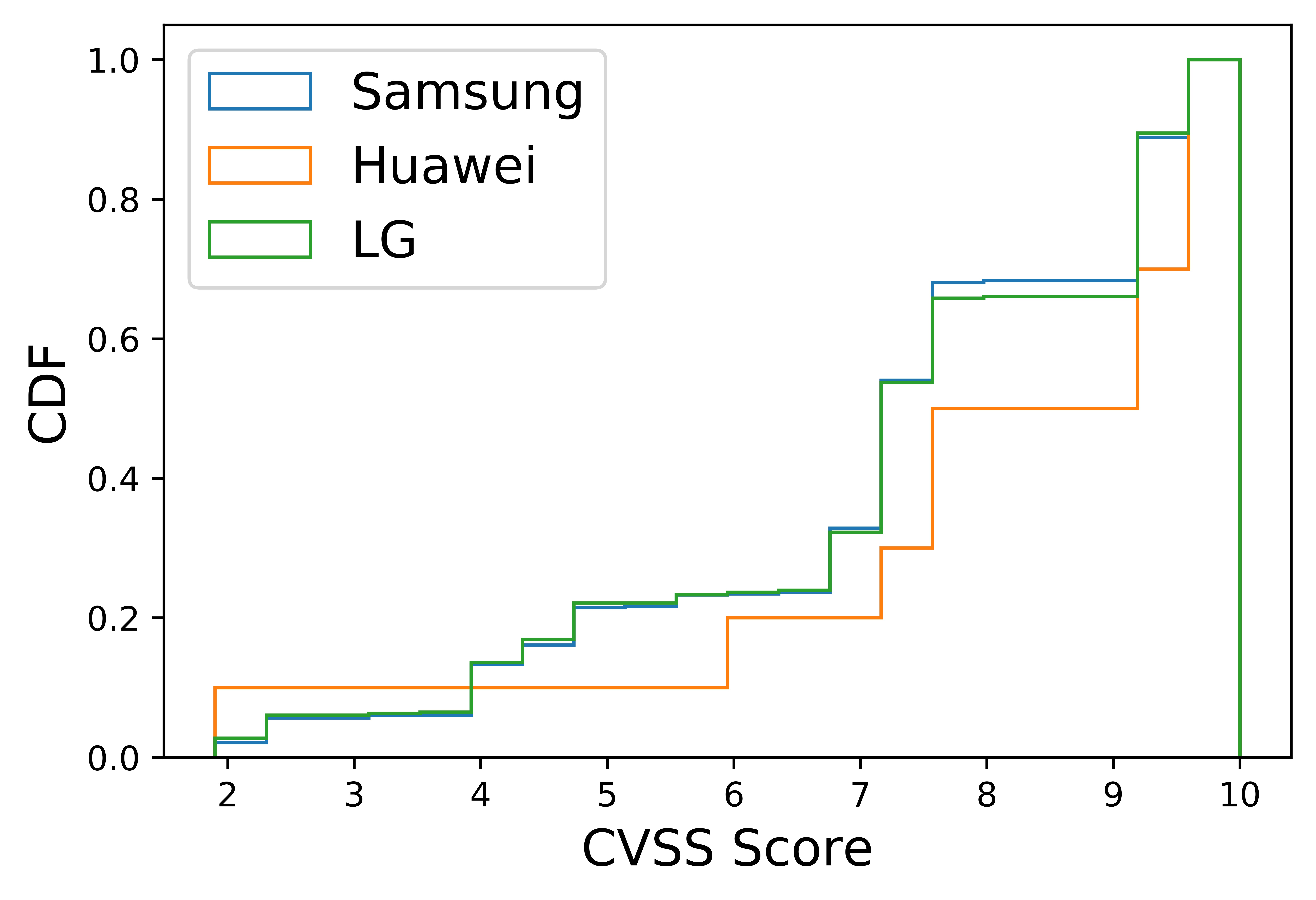}
         \caption{\footnotesize  Delay from Android}
         \label{fig:cvss_delay}
     \end{subfigure}
     \hfill
     \begin{subfigure}[b]{0.15\textwidth}
         \centering
         \includegraphics[width=\textwidth]{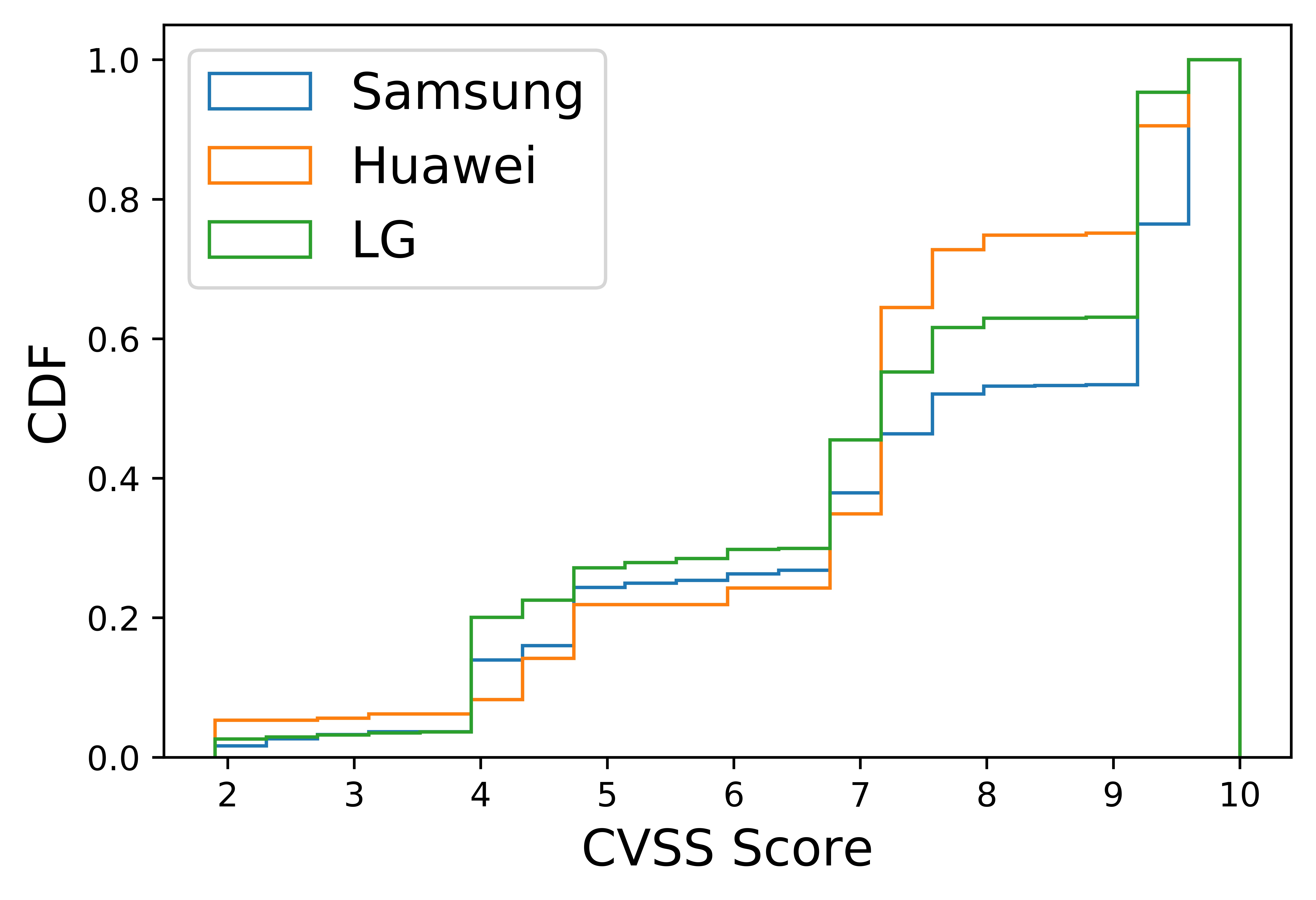}
         \caption{\footnotesize  No delay}
         \label{fig:cvss_NO}
     \end{subfigure}
     \hfill
     \begin{subfigure}[b]{0.15\textwidth}
         \centering
         \includegraphics[width=\textwidth]{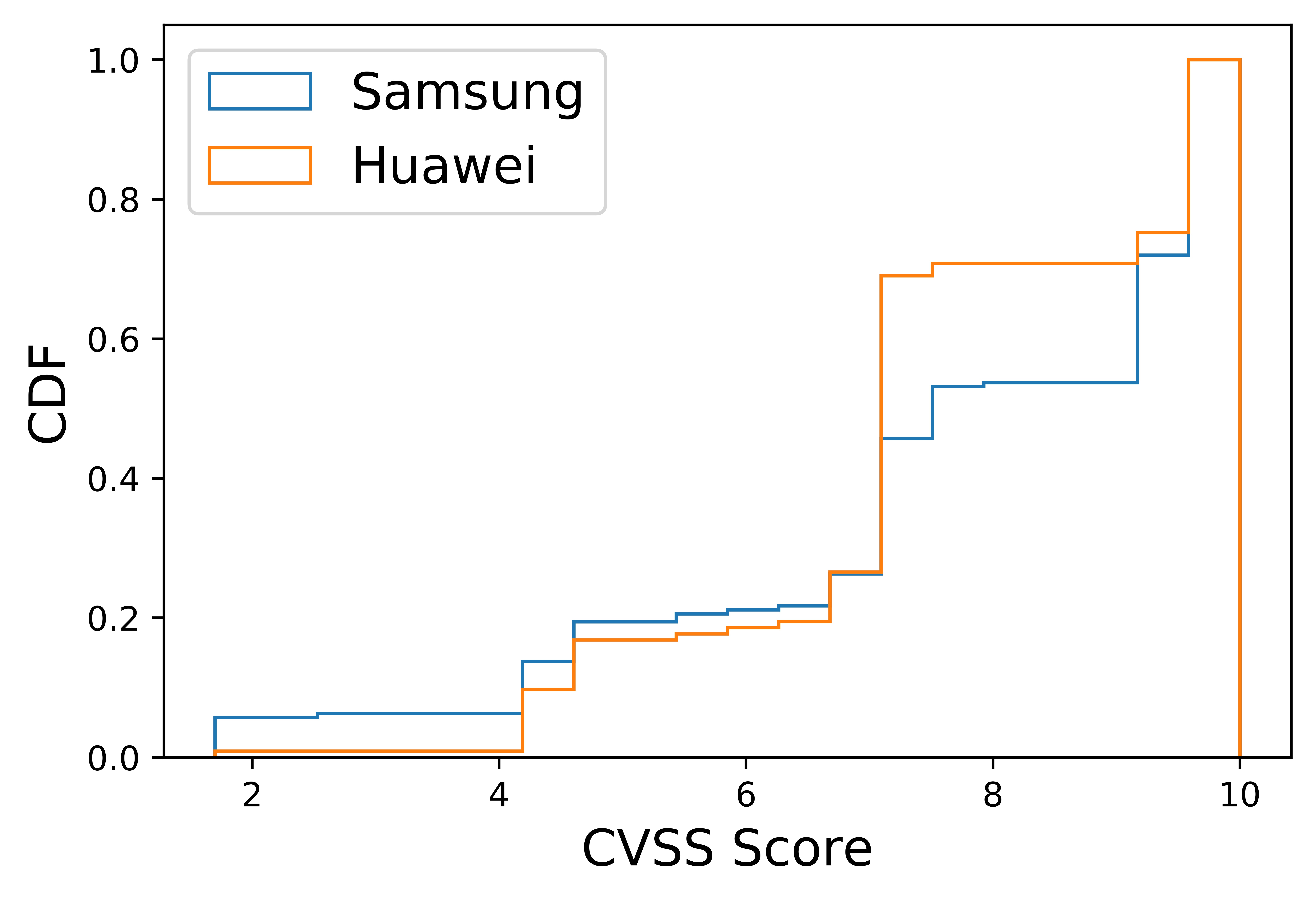}
         \caption{\footnotesize  Already included}
         \label{fig:cvss_already}
     \end{subfigure}
        \caption{Distribution of CVSS scores for different vendors. }
        \label{fig:CVSS}
\end{figure}

The time comparison of different vendors only represents one aspect of how vendors manage their security bulletins. It is also worth looking at the CVEs' severity level (CVSS score) and delay. Figure~\ref{fig:CVSS} shows a cumulative histogram of CVSS score for different vendors for CVEs with and without delay. Samsung performs better for CVEs of high CVSS scores compared to LG. When there is no delay, Samsung has more CVEs with a high CVSS score than LG. On the other hand, in the presence of delay, we are observing a higher number of CVEs with a high CVSS score in LG compared to Samsung. Furthermore, Samsung has more CVEs with a high CVSS score that have been already included compared to Huawei.  

In addition to the CVSS score, we also study the Common Weakness Enumeration (CWE), a standard for identifying the class of software weakness, of CVEs in our dataset. A CWE ID has been assigned to $92.9\%$ of our dataset. Table~\ref{tab:CWE} shows, for each vendor, the top 10 CWE IDs by the number of CVEs for three different categories, i.e., delayed CVEs, CVEs without delay, and ``already included'' CVEs. CVEs with delay in both LG and Samsung have the same top 10 CWE IDs with little changes in their order. This is mostly true for CVEs without delay in these vendors with only one difference (CWE ID 476 and CWE ID 399). But, we see a significant difference in Huawei in terms of both type and rank of CWE IDs. One of the reasons for this is that the prevalence of CWE IDs changes over time. For example, from December 2017, CWE ID 787 and CWE ID 264 correspond to around the same number of CVEs (89 and 74, respectively). Second, as we see in Figure~\ref{fig:mange}, Huawei does not mention a considerable amount of CVEs in its security bulletins (e.g., after December 2017, only $15.91\%$ of CWE ID 284 have been mentioned in Huawei security bulletins). 

\begin{table*}[]
\footnotesize{
\begin{tabular}{|c|c|c|c|c|c|c|c|c|c|}
\hline
       &                                    & \multicolumn{3}{c|}{Samsung}          & \multicolumn{3}{c|}{Huawei}         & \multicolumn{2}{c|}{LG} \\ \cline{3-10}
CWE ID & Weakness Summary                   & Delay   & WO Delay & Already Included & Delay & WO Delay & Already Included & Delay      & WO Delay   \\ \hline\hline
264    & Access Control Error               & 233 (1) & 213 (1)  & 37 (2)           & 4 (1) & 36 (4)   & 9 (5)            & 336 (1)    & 191 (1)    \\ \hline
119    & Buffer Overflow                    & 116 (2) & 185 (2)  & 41 (1)           & 3 (2) & 43 (3)   & 35 (1)           & 168 (2)    & 70 (4)     \\ \hline
200    & Information Disclosure             & 94 (3)  & 100 (3)  & 11 (5)           & -     & 16 (7)   & 5 (7)            & 124 (3)    & 88 (2)     \\ \hline
284    & Improper Access Conrol             & 49 (4)  & 89 (4)   & 12 (4)           & -     & -        & 3 (9)            & 69 (4)     & 83 (3)     \\ \hline
20     & Improper Input Validation          & 42 (5)  & 81 (5)   & 25 (3)           & -     & 23 (6)   & 10 (3)           & 62 (6)     & 50 (6)     \\ \hline
416    & Use After Free                     & 38 (6)  & 19 (10)  & 6 (8)            & 1 (3) & 27 (5)   & 10 (3)           & 66 (5)     & 15 (8)     \\ \hline
125    & Out of Band Reads                  & 30 (7)  & 48 (7)   & -                & -     & 61 (2)   & 8 (6)            & 40 (8)     & 48 (7)     \\ \hline
190    & Integer Overflow or Wraparound     & 25 (8)  & 25 (8)   & 8 (7)            & 1 (3) & 10 (8)   & 12 (2)           & 42 (7)     & 8 (9)      \\ \hline
362    & Race Condition                     & 22 (9)  & -        & -                & -     & -        & -                & 29 (9)     & -          \\ \hline
787    & Out of Band Writes                 & 12 (10) & 70 (6)   & -                & -     & 69 (1)   & -                & 17 (10)    & 68 (5)     \\ \hline
476    & Null Pointer Dereference           & -       & 24 (9)   & 9 (6)            & -     & -        & 3 (9)            & -          & -          \\ \hline
399    & Resource Management Error          & -       & -        & -                & -     & -        & -                & -          & 8 (9)      \\ \hline
400    & Uncontrolled Resource Consumption  & -       & -        & -                & -     & 7 (9)    & -                & -          & -          \\ \hline
129    & Improper Validation of Array Index & -       & -        & 5 (9)            & -     & 6 (10)   & 4 (8)            & -          & -          \\ \hline
191    & Integer Underflow                  & -       & -        & -                & -     & -        & 3 (9)            & -          & -          \\ \hline
285    & Improper Authorization             & -       & -        & 3 (10)           & -     & -        & -                & -          & -          \\ \hline
\end{tabular}}
\caption{Top 10 CWE software weaknesses by the number of CVEs. WO denotes without delay. The number in parenthesis shows that rank of a CWE on the corresponding category.}
\label{tab:CWE}
\end{table*}

\section{Discussion}
\label{sec:Discuss}

Our work sheds light at the security practices of different vendors in the Android ecosystem and it provides an important first step for better security policy recommendations. Here, we summarize some important takeaways from our study.

\subsection{Security Bulletins Structure}
\label{sub:bulletins}

Consider a case in which a CVE has been mentioned in the Android security bulletins, but it has not appeared in a vendor's security bulletins. A security practitioner cannot infer whether this CVE applies to the vendor's Android device solely by  looking at the vendor's security bulletins due to the possibility of delay in the CVE announcement and the possibility that the vendor's patch has already included that CVE. As a result, we suggest that all vendors follow Samsung and have a section that mentions which CVEs are \textit{not applicable} to their devices. Based on Section~\ref{sub:Parts}, we do not have a bulletin reference for already included CVEs. It might be better if a vendor can provide that information, enabling security practitioners to associate CVEs to a patch correctly. 

\subsection{Public Repositories vs. SVE and LVE}
\label{sub:SVE&LVE}
As mentioned in Section~\ref{sec:Data}, Samsung has SVE and LG has LVE. The number of SVEs and LVEs are 295 and 63, respectively. For SVEs, Samsung only provides severity, affected versions, reported date, disclosure status, and description. For LVEs, LG provides the same except the disclosure status. Both of these vendors do not provide any corresponding CVE number for these SVEs and LVEs. However, we have found the corresponding CVEs for some SVEs and LVEs by manual search. But even with manual search, we cannot find any corresponding CVEs for some of them. CVEDetails and NVD database have been established via community effort to provide a comprehensive database for anyone to facilitate vulnerability mitigation. As a result, we suggest that both Samsung and LG provide CVEs for their corresponding SVEs and LVEs.

\subsection{Inconsistency}
\label{sub:inconsistent}
In our previous work~\cite{farhang2019hey}, we observed various inconsistencies in Android security bulletins and CVEDetails. Further, there have been recent efforts to find inconsistencies in public security reports using natural language processing~\cite{dong2019towards}. However, here, we observe inconsistencies within a vendor. In Samsung, we observe an inconsistency in terms of both severity level and whether a CVE is applicable to Samsung devices. Note that these CVEs are mentioned only once in Android security bulletins. For example, \texttt{CVE-2016-5342} mentioned in Nov. 2016 with a high severity level is mentioned again in July 2018 with a moderate severity level. Moreover, \texttt{CVE-2014-9981} is mentioned in Oct. 2017 as not applicable to Samsung devices. However, it again appears in February 2018 in the \textit{in addition} part. A vendor's security bulletin is a reference point for a security professional to check vulnerabilities. If there exists inconsistency, this leaves a security professional with uncertainty. Therefore, in addition to consistency among public repositories, consistency within a vendor's security bulletin is essential.

\section{Related Work}
\label{sec:Related}

\textbf{Android Security and Software Updates.} Research efforts on the security of Android are immense and include a wide spectrum (vulnerability finding, attack investigation, and secure infrastructure)~\cite{vidas2011all, xing2014upgrading, enck2009understanding, shabtai2010securing, thomas2015security}. Similar to other software, Android vendors maintain the security of their devices by developing and issuing patches regularly. Nappa et al.~\cite{nappa2015attack} studied the vulnerabilities life cycle in client applications, and Li and Paxson~\cite{li2017large} investigated the patch development life cycle in open source software projects. The issues of automatic updates and semi-automatic updates have been investigated in~\cite{edwards2008security} and \cite{mathur2017impact}, respectively. Farhang et al.~\cite{farhang2018take} differentiate between update and upgrade and study the upgrade practices in the Windows operating system. 
Contrary to previous work, we investigate the security practices of different vendors in terms of vulnerability and patch management by gathering and analyzing different vendors' security bulletins.

\textbf{Security Reports.} Vulnerability reports and security bulletins have been studied in different domains. In 2016, the U.S. Federal Trade Commission (FTC) started to study major mobile device vendors' security update practices~\cite{FTC_Study,us2018mobile}. 
Arora et al.~\cite{arora2010empirical} showed that disclosure accelerates patch release.
Bugs remediation can be improved by interaction between software developers and bug reporters~\cite{breu2010information}.
Security bugs that have been found by reputable professionals get patched faster~\cite{guo2010characterizing}.
Missing information in vulnerability reports endangers vulnerability reproduction~\cite{mu2018understanding}.
Most closely to our work are~\cite{linares2017empirical} and \cite{farhang2019hey}. Linares-V{\'a}squez et al.~\cite{linares2017empirical} studied CWE hierarchies, vulnerability types, Android layers affected by vulnerabilities by collecting 660 Android-related vulnerabilities from Android security bulletins. Farhang et al.~\cite{farhang2019hey} not only conducted a similar study with a more comprehensive dataset (2,470 Android-related vulnerabilities), but also investigated new aspects like patching vulnerabilities originating from Qualcomm and Linux. Here, we study the commonalities and differences among vendors in the Android ecosystem by collecting Android-related vulnerabilities from different vendors, i.e., Samsung, LG, and Huawei. 

\section{Conclusion}
\label{sec:Conclusion}

We provided a comprehensive study of multiple Android vendors' security bulletins to better understand how different vendors manage their security bulletins. By collecting 3,171 unique CVEs from Android, Samsung, LG, and Huawei security bulletins, we investigated bulletin management, delay related to Android stack layers, and CWE ID in different layers. We found that (i) vendors have different structures for vulnerability reporting, (ii) Qualcomm-related CVEs and the rest of external layers' CVEs are handled differently by vendors, (iii) the likelihood of delay for CVEs with Android Git repository is low. 
\balance

\textbf{Acknowledgments:} We thank the reviewers for their insightful comments and suggestions. This work was supported in part by the National Science Foundation under Grant CNS-1850510.

\clearpage

\bibliographystyle{ACM-Reference-Format}
\bibliography{sample-base}


\begin{thebibliography}{37}


\ifx \showCODEN    \undefined \def \showCODEN     #1{\unskip}     \fi
\ifx \showDOI      \undefined \def \showDOI       #1{#1}\fi
\ifx \showISBNx    \undefined \def \showISBNx     #1{\unskip}     \fi
\ifx \showISBNxiii \undefined \def \showISBNxiii  #1{\unskip}     \fi
\ifx \showISSN     \undefined \def \showISSN      #1{\unskip}     \fi
\ifx \showLCCN     \undefined \def \showLCCN      #1{\unskip}     \fi
\ifx \shownote     \undefined \def \shownote      #1{#1}          \fi
\ifx \showarticletitle \undefined \def \showarticletitle #1{#1}   \fi
\ifx \showURL      \undefined \def \showURL       {\relax}        \fi
\providecommand\bibfield[2]{#2}
\providecommand\bibinfo[2]{#2}
\providecommand\natexlab[1]{#1}
\providecommand\showeprint[2][]{arXiv:#2}

\bibitem[\protect\citeauthoryear{Aafer, Zhang, and Du}{Aafer
  et~al\mbox{.}}{2016}]%
        {aafer2016harvesting}
\bibfield{author}{\bibinfo{person}{Yousra Aafer}, \bibinfo{person}{Xiao Zhang},
  {and} \bibinfo{person}{Wenliang Du}.} \bibinfo{year}{2016}\natexlab{}.
\newblock \showarticletitle{Harvesting inconsistent security configurations in
  custom {A}ndroid roms via differential analysis}. In
  \bibinfo{booktitle}{\emph{Proceedings of the 25th USENIX Security Symposium
  (USENIX Security)}}. \bibinfo{pages}{1153--1168}.
\newblock


\bibitem[\protect\citeauthoryear{{AOSP}}{{AOSP}}{[n. d.]}]%
        {Android_Bulletin}
\bibfield{author}{\bibinfo{person}{{AOSP}}.} \bibinfo{year}{[n.
  d.]}\natexlab{}.
\newblock \bibinfo{title}{Android Security Bulletin}.
\newblock
  \bibinfo{howpublished}{\url{https://source.android.com/security/bulletin}}.
\newblock
\newblock
\shownote{Accessed: 10/13/2019.}


\bibitem[\protect\citeauthoryear{{AOSP}}{{AOSP}}{2015a}]%
        {Android_Security_Bulletin_2015_08}
\bibfield{author}{\bibinfo{person}{{AOSP}}.} \bibinfo{year}{2015}\natexlab{a}.
\newblock \bibinfo{title}{Android Security Bulletin 2015-08}.
\newblock
  \bibinfo{howpublished}{\url{https://source.android.com/security/bulletin/2015-08-01}}.
\newblock
\newblock
\shownote{Accessed: 01/30/2019.}


\bibitem[\protect\citeauthoryear{{AOSP}}{{AOSP}}{2015b}]%
        {Android_Security_Bulletin_2015_12}
\bibfield{author}{\bibinfo{person}{{AOSP}}.} \bibinfo{year}{2015}\natexlab{b}.
\newblock \bibinfo{title}{Android Security Bulletin 2015-12}.
\newblock
  \bibinfo{howpublished}{\url{https://source.android.com/security/bulletin/2015-12-01}}.
\newblock
\newblock
\shownote{Accessed: 01/26/2019.}


\bibitem[\protect\citeauthoryear{Arora, Krishnan, Telang, and Yang}{Arora
  et~al\mbox{.}}{2010}]%
        {arora2010empirical}
\bibfield{author}{\bibinfo{person}{Ashish Arora}, \bibinfo{person}{Ramayya
  Krishnan}, \bibinfo{person}{Rahul Telang}, {and} \bibinfo{person}{Yubao
  Yang}.} \bibinfo{year}{2010}\natexlab{}.
\newblock \showarticletitle{An empirical analysis of software vendors' patch
  release behavior: {I}mpact of vulnerability disclosure}.
\newblock \bibinfo{journal}{\emph{Information Systems Research}}
  \bibinfo{volume}{21}, \bibinfo{number}{1} (\bibinfo{year}{2010}),
  \bibinfo{pages}{115--132}.
\newblock


\bibitem[\protect\citeauthoryear{Breu, Premraj, Sillito, and Zimmermann}{Breu
  et~al\mbox{.}}{2010}]%
        {breu2010information}
\bibfield{author}{\bibinfo{person}{Silvia Breu}, \bibinfo{person}{Rahul
  Premraj}, \bibinfo{person}{Jonathan Sillito}, {and} \bibinfo{person}{Thomas
  Zimmermann}.} \bibinfo{year}{2010}\natexlab{}.
\newblock \showarticletitle{Information needs in bug reports: Improving
  cooperation between developers and users}. In
  \bibinfo{booktitle}{\emph{Proceedings of the 2010 ACM Conference on Computer
  Supported Cooperative Work (CSCW)}}. ACM, \bibinfo{pages}{301--310}.
\newblock


\bibitem[\protect\citeauthoryear{Commission}{Commission}{2016}]%
        {FTC_Study}
\bibfield{author}{\bibinfo{person}{U.S. Federal~Trade Commission}.}
  \bibinfo{year}{2016}\natexlab{}.
\newblock \bibinfo{title}{FTC to study mobile device industry's security update
  practices}.
\newblock
  \bibinfo{howpublished}{\url{https://www.ftc.gov/news-events/press-releases/2016/05/ftc-study-mobile-device-industrys-security-update-practices}}.
\newblock
\newblock
\shownote{Accessed: 02/02/2020.}


\bibitem[\protect\citeauthoryear{Commission et~al\mbox{.}}{Commission
  et~al\mbox{.}}{2018}]%
        {us2018mobile}
\bibfield{author}{\bibinfo{person}{U.S. Federal~Trade Commission}
  {et~al\mbox{.}}} \bibinfo{year}{2018}\natexlab{}.
\newblock \bibinfo{title}{Mobile security updates: Understanding the issues}.
\newblock
  \bibinfo{howpublished}{\url{https://www.ftc.gov/system/files/documents/reports/mobile-security-updates-understanding-issues/mobile_security_updates_understanding_the_issues_publication_final.pdf}}.
\newblock
\newblock
\shownote{Accessed: 02/02/2020.}


\bibitem[\protect\citeauthoryear{Dong, Guo, Chen, Xing, Zhang, and Wang}{Dong
  et~al\mbox{.}}{2019}]%
        {dong2019towards}
\bibfield{author}{\bibinfo{person}{Ying Dong}, \bibinfo{person}{Wenbo Guo},
  \bibinfo{person}{Yueqi Chen}, \bibinfo{person}{Xinyu Xing},
  \bibinfo{person}{Yuqing Zhang}, {and} \bibinfo{person}{Gang Wang}.}
  \bibinfo{year}{2019}\natexlab{}.
\newblock \showarticletitle{Towards the detection of inconsistencies in public
  security vulnerability reports}. In \bibinfo{booktitle}{\emph{Proceedings of
  the 28th USENIX Security Symposium (USENIX Security)}}.
  \bibinfo{pages}{869--885}.
\newblock


\bibitem[\protect\citeauthoryear{Edwards, Poole, and Stoll}{Edwards
  et~al\mbox{.}}{2008}]%
        {edwards2008security}
\bibfield{author}{\bibinfo{person}{Keith Edwards},
  \bibinfo{person}{Erika~Shehan Poole}, {and} \bibinfo{person}{Jennifer
  Stoll}.} \bibinfo{year}{2008}\natexlab{}.
\newblock \showarticletitle{Security automation considered harmful?}. In
  \bibinfo{booktitle}{\emph{Proceedings of the 2007 Workshop on New Security
  Paradigms (NSPW)}}. ACM, \bibinfo{pages}{33--42}.
\newblock


\bibitem[\protect\citeauthoryear{Enck, Ongtang, and McDaniel}{Enck
  et~al\mbox{.}}{2009}]%
        {enck2009understanding}
\bibfield{author}{\bibinfo{person}{William Enck}, \bibinfo{person}{Machigar
  Ongtang}, {and} \bibinfo{person}{Patrick McDaniel}.}
  \bibinfo{year}{2009}\natexlab{}.
\newblock \showarticletitle{Understanding {A}ndroid security}.
\newblock \bibinfo{journal}{\emph{IEEE Security \& Privacy}}
  \bibinfo{number}{1} (\bibinfo{year}{2009}), \bibinfo{pages}{50--57}.
\newblock


\bibitem[\protect\citeauthoryear{Farhang, Kirdan, Laszka, and
  Grossklags}{Farhang et~al\mbox{.}}{2019}]%
        {farhang2019hey}
\bibfield{author}{\bibinfo{person}{Sadegh Farhang},
  \bibinfo{person}{Mehmet~Bahadir Kirdan}, \bibinfo{person}{Aron Laszka}, {and}
  \bibinfo{person}{Jens Grossklags}.} \bibinfo{year}{2019}\natexlab{}.
\newblock \showarticletitle{Hey Google, What Exactly Do Your Security Patches
  Tell Us? A Large-Scale Empirical Study on Android Patched Vulnerabilities}.
\newblock \bibinfo{journal}{\emph{arXiv preprint arXiv:1905.09352}}
  (\bibinfo{year}{2019}).
\newblock


\bibitem[\protect\citeauthoryear{Farhang, Laszka, and Grossklags}{Farhang
  et~al\mbox{.}}{2018a}]%
        {farhang2018economic}
\bibfield{author}{\bibinfo{person}{Sadegh Farhang}, \bibinfo{person}{Aron
  Laszka}, {and} \bibinfo{person}{Jens Grossklags}.}
  \bibinfo{year}{2018}\natexlab{a}.
\newblock \showarticletitle{An economic study of the effect of {A}ndroid
  platform fragmentation on security updates}. In
  \bibinfo{booktitle}{\emph{Proceedings of the 22nd International Conference on
  Financial Cryptography and Data Security (FC)}}. Springer,
  \bibinfo{pages}{119--137}.
\newblock


\bibitem[\protect\citeauthoryear{Farhang, Weidman, Kamani, Grossklags, and
  Liu}{Farhang et~al\mbox{.}}{2018b}]%
        {farhang2018take}
\bibfield{author}{\bibinfo{person}{Sadegh Farhang}, \bibinfo{person}{Jake
  Weidman}, \bibinfo{person}{Mohammad~Mahdi Kamani}, \bibinfo{person}{Jens
  Grossklags}, {and} \bibinfo{person}{Peng Liu}.}
  \bibinfo{year}{2018}\natexlab{b}.
\newblock \showarticletitle{Take It or Leave It: {A} Survey Study on Operating
  System Upgrade Practices}. In \bibinfo{booktitle}{\emph{Proceedings of the
  34th Annual Computer Security Applications Conference (ACSAC)}}. ACM,
  \bibinfo{pages}{490--504}.
\newblock


\bibitem[\protect\citeauthoryear{{Google}}{{Google}}{[n. d.]}]%
        {google_aosp_git}
\bibfield{author}{\bibinfo{person}{{Google}}.} \bibinfo{year}{[n.
  d.]}\natexlab{}.
\newblock \bibinfo{title}{AOSP Git}.
\newblock \bibinfo{howpublished}{\url{https://android.googlesource.com/}}.
\newblock
\newblock
\shownote{Accessed: 10/5/2019.}


\bibitem[\protect\citeauthoryear{Guo, Zimmermann, Nagappan, and Murphy}{Guo
  et~al\mbox{.}}{2010}]%
        {guo2010characterizing}
\bibfield{author}{\bibinfo{person}{Philip Guo}, \bibinfo{person}{Thomas
  Zimmermann}, \bibinfo{person}{Nachiappan Nagappan}, {and}
  \bibinfo{person}{Brendan Murphy}.} \bibinfo{year}{2010}\natexlab{}.
\newblock \showarticletitle{Characterizing and predicting which bugs get fixed:
  {A}n empirical study of {M}icrosoft {W}indows}. In
  \bibinfo{booktitle}{\emph{Proceedings of the 32nd ACM/IEEE International
  Conference on Software Engineering (ICSE)}}. ACM, \bibinfo{pages}{495--504}.
\newblock


\bibitem[\protect\citeauthoryear{{Huawei}}{{Huawei}}{[n. d.]a}]%
        {Huawei_Bulletin}
\bibfield{author}{\bibinfo{person}{{Huawei}}.} \bibinfo{year}{[n.
  d.]}\natexlab{a}.
\newblock \bibinfo{title}{Android Security Bulletin}.
\newblock
  \bibinfo{howpublished}{\url{https://consumer.huawei.com/en/support/bulletin/}}.
\newblock
\newblock
\shownote{Accessed: 10/13/2019.}


\bibitem[\protect\citeauthoryear{{Huawei}}{{Huawei}}{[n. d.]b}]%
        {huawei_emui}
\bibfield{author}{\bibinfo{person}{{Huawei}}.} \bibinfo{year}{[n.
  d.]}\natexlab{b}.
\newblock \bibinfo{title}{{EMUI/Magic UI}}.
\newblock
  \bibinfo{howpublished}{\url{https://consumer.huawei.com/en/support/bulletin/}}.
\newblock
\newblock
\shownote{Accessed: 10/5/2019.}


\bibitem[\protect\citeauthoryear{{Huawei}}{{Huawei}}{[n. d.]c}]%
        {huawei_security_advisory}
\bibfield{author}{\bibinfo{person}{{Huawei}}.} \bibinfo{year}{[n.
  d.]}\natexlab{c}.
\newblock \bibinfo{title}{{Security Advisories}}.
\newblock
  \bibinfo{howpublished}{\url{https://www.huawei.com/en/psirt/all-bulletins}}.
\newblock
\newblock
\shownote{Accessed: 10/5/2019.}


\bibitem[\protect\citeauthoryear{{IDC}}{{IDC}}{[n. d.]}]%
        {Android_Market_Share}
\bibfield{author}{\bibinfo{person}{{IDC}}.} \bibinfo{year}{[n. d.]}\natexlab{}.
\newblock \bibinfo{title}{Android Market Share}.
\newblock
  \bibinfo{howpublished}{\url{https://www.idc.com/promo/smartphone-market-share/os}}.
\newblock
\newblock
\shownote{Accessed: 10/13/2019.}


\bibitem[\protect\citeauthoryear{{LG}}{{LG}}{[n. d.]}]%
        {LG_Bulletin}
\bibfield{author}{\bibinfo{person}{{LG}}.} \bibinfo{year}{[n. d.]}\natexlab{}.
\newblock \bibinfo{title}{Android Security Bulletin}.
\newblock
  \bibinfo{howpublished}{\url{https://lgsecurity.lge.com/security_updates_mobile.html}}.
\newblock
\newblock
\shownote{Accessed: 10/13/2019.}


\bibitem[\protect\citeauthoryear{Li and Paxson}{Li and Paxson}{2017}]%
        {li2017large}
\bibfield{author}{\bibinfo{person}{Frank Li} {and} \bibinfo{person}{Vern
  Paxson}.} \bibinfo{year}{2017}\natexlab{}.
\newblock \showarticletitle{A large-scale empirical study of security patches}.
  In \bibinfo{booktitle}{\emph{Proceedings of the 24th ACM SIGSAC Conference on
  Computer and Communications Security (CCS)}}. ACM,
  \bibinfo{pages}{2201--2215}.
\newblock


\bibitem[\protect\citeauthoryear{Linares-V{\'a}squez, Bavota, and
  Escobar-Vel{\'a}squez}{Linares-V{\'a}squez et~al\mbox{.}}{2017}]%
        {linares2017empirical}
\bibfield{author}{\bibinfo{person}{Mario Linares-V{\'a}squez},
  \bibinfo{person}{Gabriele Bavota}, {and} \bibinfo{person}{Camilo
  Escobar-Vel{\'a}squez}.} \bibinfo{year}{2017}\natexlab{}.
\newblock \showarticletitle{An empirical study on {A}ndroid-related
  vulnerabilities}. In \bibinfo{booktitle}{\emph{IEEE/ACM 14th International
  Conference on Mining Software Repositories (MSR)}}. IEEE,
  \bibinfo{pages}{2--13}.
\newblock


\bibitem[\protect\citeauthoryear{Mathur and Chetty}{Mathur and Chetty}{2017}]%
        {mathur2017impact}
\bibfield{author}{\bibinfo{person}{Arunesh Mathur} {and}
  \bibinfo{person}{Marshini Chetty}.} \bibinfo{year}{2017}\natexlab{}.
\newblock \showarticletitle{Impact of user characteristics on attitudes towards
  automatic mobile application updates}. In
  \bibinfo{booktitle}{\emph{Proceedings of the 13th Symposium on Usable Privacy
  and Security (SOUPS)}}.
\newblock


\bibitem[\protect\citeauthoryear{{MongoDB}}{{MongoDB}}{[n. d.]}]%
        {MongoDB}
\bibfield{author}{\bibinfo{person}{{MongoDB}}.} \bibinfo{year}{[n.
  d.]}\natexlab{}.
\newblock \bibinfo{title}{What is {MongoDB}?}
\newblock
  \bibinfo{howpublished}{\url{https://www.mongodb.com/what-is-mongodb}}.
\newblock
\newblock
\shownote{Accessed: 01/23/2019.}


\bibitem[\protect\citeauthoryear{Mu, Cuevas, Yang, Hu, Xing, Mao, and Wang}{Mu
  et~al\mbox{.}}{2018}]%
        {mu2018understanding}
\bibfield{author}{\bibinfo{person}{Dongliang Mu}, \bibinfo{person}{Alejandro
  Cuevas}, \bibinfo{person}{Limin Yang}, \bibinfo{person}{Hang Hu},
  \bibinfo{person}{Xinyu Xing}, \bibinfo{person}{Bing Mao}, {and}
  \bibinfo{person}{Gang Wang}.} \bibinfo{year}{2018}\natexlab{}.
\newblock \showarticletitle{Understanding the reproducibility of crowd-reported
  security vulnerabilities}. In \bibinfo{booktitle}{\emph{Proceedings of the
  27th USENIX Security Symposium (USENIX Security)}}.
  \bibinfo{pages}{919--936}.
\newblock


\bibitem[\protect\citeauthoryear{Nappa, Johnson, Bilge, Caballero, and
  Dumitras}{Nappa et~al\mbox{.}}{2015}]%
        {nappa2015attack}
\bibfield{author}{\bibinfo{person}{Antonio Nappa}, \bibinfo{person}{Richard
  Johnson}, \bibinfo{person}{Leyla Bilge}, \bibinfo{person}{Juan Caballero},
  {and} \bibinfo{person}{Tudor Dumitras}.} \bibinfo{year}{2015}\natexlab{}.
\newblock \showarticletitle{The attack of the clones: {A} study of the impact
  of shared code on vulnerability patching}. In
  \bibinfo{booktitle}{\emph{Proceedings of the IEEE Symposium on Security and
  Privacy (S\&P)}}. IEEE, \bibinfo{pages}{692--708}.
\newblock


\bibitem[\protect\citeauthoryear{\"Ozkan}{\"Ozkan}{[n. d.]}]%
        {Cve_Details}
\bibfield{author}{\bibinfo{person}{Serkan \"Ozkan}.} \bibinfo{year}{[n.
  d.]}\natexlab{}.
\newblock \bibinfo{title}{{CVE} Details}.
\newblock \bibinfo{howpublished}{\url{https://www.cvedetails.com}}.
\newblock
\newblock
\shownote{Accessed: 01/23/2019.}


\bibitem[\protect\citeauthoryear{{Samsung}}{{Samsung}}{[n. d.]}]%
        {Samsung_Bulletin}
\bibfield{author}{\bibinfo{person}{{Samsung}}.} \bibinfo{year}{[n.
  d.]}\natexlab{}.
\newblock \bibinfo{title}{Android Security Bulletin}.
\newblock
  \bibinfo{howpublished}{\url{https://security.samsungmobile.com/securityUpdate.smsb}}.
\newblock
\newblock
\shownote{Accessed: 10/13/2019.}


\bibitem[\protect\citeauthoryear{{Selenium Project}}{{Selenium Project}}{[n.
  d.]}]%
        {Selenium}
\bibfield{author}{\bibinfo{person}{{Selenium Project}}.} \bibinfo{year}{[n.
  d.]}\natexlab{}.
\newblock \bibinfo{title}{Selenium}.
\newblock \bibinfo{howpublished}{\url{https://www.seleniumhq.org}}.
\newblock
\newblock
\shownote{Accessed: 01/23/2019.}


\bibitem[\protect\citeauthoryear{Shabtai, Fledel, and Elovici}{Shabtai
  et~al\mbox{.}}{2010}]%
        {shabtai2010securing}
\bibfield{author}{\bibinfo{person}{Asaf Shabtai}, \bibinfo{person}{Yuval
  Fledel}, {and} \bibinfo{person}{Yuval Elovici}.}
  \bibinfo{year}{2010}\natexlab{}.
\newblock \showarticletitle{Securing {A}ndroid-powered mobile devices using
  {SEL}inux}.
\newblock \bibinfo{journal}{\emph{IEEE Security \& Privacy}}
  \bibinfo{volume}{8}, \bibinfo{number}{3} (\bibinfo{year}{2010}),
  \bibinfo{pages}{36--44}.
\newblock


\bibitem[\protect\citeauthoryear{{statcounter}}{{statcounter}}{2019}]%
        {Device_Market_Share}
\bibfield{author}{\bibinfo{person}{{statcounter}}.}
  \bibinfo{year}{2019}\natexlab{}.
\newblock \bibinfo{title}{Mobile Device Market Share}.
\newblock
  \bibinfo{howpublished}{\url{https://gs.statcounter.com/vendor-market-share/mobile/worldwide/2019}}.
\newblock
\newblock
\shownote{Accessed: 10/13/2019.}


\bibitem[\protect\citeauthoryear{Thomas, Beresford, and Rice}{Thomas
  et~al\mbox{.}}{2015}]%
        {thomas2015security}
\bibfield{author}{\bibinfo{person}{Daniel Thomas}, \bibinfo{person}{Alastair
  Beresford}, {and} \bibinfo{person}{Andrew Rice}.}
  \bibinfo{year}{2015}\natexlab{}.
\newblock \showarticletitle{Security metrics for the {A}ndroid ecosystem}. In
  \bibinfo{booktitle}{\emph{Proceedings of the 5th Annual ACM CCS Workshop on
  Security and Privacy in Smartphones and Mobile Devices}}.
  \bibinfo{pages}{87--98}.
\newblock


\bibitem[\protect\citeauthoryear{Vidas, Votipka, and Christin}{Vidas
  et~al\mbox{.}}{2011}]%
        {vidas2011all}
\bibfield{author}{\bibinfo{person}{Timothy Vidas}, \bibinfo{person}{Daniel
  Votipka}, {and} \bibinfo{person}{Nicolas Christin}.}
  \bibinfo{year}{2011}\natexlab{}.
\newblock \showarticletitle{All your droid are belong to us: {A} survey of
  current {A}ndroid attacks}. In \bibinfo{booktitle}{\emph{Proceedings of the
  5th USENIX Workshop on Offensive Technologies (WOOT)}}.
  \bibinfo{pages}{81--90}.
\newblock


\bibitem[\protect\citeauthoryear{{Wikipedia}}{{Wikipedia}}{[n. d.]}]%
        {JSON}
\bibfield{author}{\bibinfo{person}{{Wikipedia}}.} \bibinfo{year}{[n.
  d.]}\natexlab{}.
\newblock \bibinfo{title}{JSON}.
\newblock \bibinfo{howpublished}{\url{https://en.wikipedia.org/wiki/JSON}}.
\newblock
\newblock
\shownote{Accessed: 01/23/2019.}


\bibitem[\protect\citeauthoryear{Wu, Grace, Zhou, Wu, and Jiang}{Wu
  et~al\mbox{.}}{2013}]%
        {wu2013impact}
\bibfield{author}{\bibinfo{person}{Lei Wu}, \bibinfo{person}{Michael Grace},
  \bibinfo{person}{Yajin Zhou}, \bibinfo{person}{Chiachih Wu}, {and}
  \bibinfo{person}{Xuxian Jiang}.} \bibinfo{year}{2013}\natexlab{}.
\newblock \showarticletitle{The impact of vendor customizations on {A}ndroid
  security}. In \bibinfo{booktitle}{\emph{Proceedings of the 2013 ACM SIGSAC
  Conference on Computer \& Communications Security (CCS)}}. ACM,
  \bibinfo{pages}{623--634}.
\newblock


\bibitem[\protect\citeauthoryear{Xing, Pan, Wang, Yuan, and Wang}{Xing
  et~al\mbox{.}}{2014}]%
        {xing2014upgrading}
\bibfield{author}{\bibinfo{person}{Luyi Xing}, \bibinfo{person}{Xiaorui Pan},
  \bibinfo{person}{Rui Wang}, \bibinfo{person}{Kan Yuan}, {and}
  \bibinfo{person}{XiaoFeng Wang}.} \bibinfo{year}{2014}\natexlab{}.
\newblock \showarticletitle{Upgrading your {A}ndroid, elevating my malware:
  {P}rivilege escalation through mobile {OS} updating}. In
  \bibinfo{booktitle}{\emph{Proceedings of the IEEE Symposium on Security and
  Privacy (S\&P)}}. IEEE, \bibinfo{pages}{393--408}.
\newblock


\end{thebibliography}




\end{document}